\documentclass[conference]{IEEEtran}
\IEEEoverridecommandlockouts
        
\usepackage{cite}
\usepackage{amsmath,amssymb,amsfonts}
\usepackage{algorithmic}
\usepackage{graphicx}
\usepackage{textcomp}
\usepackage{xcolor}
\usepackage{booktabs}
\usepackage{multirow}
\def\BibTeX{{\rm B\kern-.05em{\sc i\kern-.025em b}\kern-.08em
    T\kern-.1667em\lower.7ex\hbox{E}\kern-.125emX}}
\begin{document}

\title{The Development of Reflective Practice on a Work-Based Software Engineering Program: A Longitudinal Study}

\author{\IEEEauthorblockN{Matthew Barr}
\IEEEauthorblockA{\textit{School of Computing Science} \\
\textit{University of Glasgow}\\
Glasgow, UK \\
Matthew.Barr@glasgow.ac.uk}
\and
\IEEEauthorblockN{Syed Waqar Nabi}
\IEEEauthorblockA{\textit{School of Computing Science} \\
\textit{University of Glasgow}\\
Glasgow, UK \\
Syed.Nabi@glasgow.ac.uk}
\and
\IEEEauthorblockN{Oana Andrei}
\IEEEauthorblockA{\textit{School of Computing Science} \\
\textit{University of Glasgow}\\
Glasgow, UK \\
Oana.Andrei@glasgow.ac.uk}
}

\maketitle

\begin{abstract}
  This study examines the development of reflective practice among students in an undergraduate work-based Software Engineering program. Using two established models of reflection -- Boud \textit{et al}.'s Model of Reflective Process and Bain \textit{et al.}'s 5R Framework for Reflection -- we analyse a series of reflective reports submitted by students over four years. Our longitudinal analysis reveals clear trends in how students' reflective abilities evolve over the course of the program. We find that more sophisticated forms of reflection, such as integration of knowledge, appropriation of skills, and reconstruction of practice, increase markedly in prevalence in later years. The complementary nature of workplace experience and university study is highlighted in students' reflections, demonstrating a key benefit of the work-based learning approach. Our findings provide insight into how reflective practice develops in Software Engineering education and suggest potential value in incorporating more structured reflection into traditional degree programs.
\end{abstract}


\section{Introduction}

Experiential learning -- learning by doing -- is an approach to learning that can trace its roots to humanity's origins, but which has been recognized as a formal educational approach only in recent decades. A specialized form of experiential learning is Work-based Learning (WBL), where learning happens in the workplace rather than a conventional classroom. \emph{Reflection} is a central and necessary aspect of the experiential learning approach, including WBL. 

Reflection has a critical role in WBL because the way learning happens in the workplace is very different from the classroom: in the absence of a structured learning environment, learning is often unintentional and may even happen without the learner's knowledge~\cite{Boyd_Fales_1983}. While workplace experience will inevitably lead to some learning, the risk is that the proficiency remains limited to a particular kind of behaviour. Reflection helps to create and clarify meaning from workplace experiences. John Dewey's conception of how we learn~\cite{Dewey33} gives experiential learning a central role, as a means of growth that leads to a fundamental change in perspective in the learner. This growth is achieved only when the experience is accompanied by reflection~\cite{Rodgers_LaBoskey_2016}.

Raelin~\cite{Raelin_1997} offers a model that conceptualizes WBL and the role that reflection plays in this pedagogy. Raelin conceives the learning journey starting with \emph{conceptualization} in the classroom: theoretical learning and explicit knowledge. The lab environment provides an opportunity for experimentation, thus adding context. However, learning -- decoupled from any real-world context or concerns -- is still effectively theoretical at this stage. A conventional learning environment is usually limited to this point, whereas in a WBL setting, the workplace provides a practical, real-world dimension to learning, where theoretical knowledge can now be \emph{cemented} via experience. The final crucial step in this model is \emph{reflection}. Without reflection, learning is tacit, often limited to particular kinds of behaviour. Reflection creates the space to convert at least some of the tacit learning into explicit knowledge -- a change in \emph{perspective} -- which can thereon be called upon in a more deliberate manner. Raelin, for example, notes: ``...reflection is required to bring the inherent tacit knowledge of experience to the surface. It thus contributes to the reconstruction of meaning''~\cite{Raelin_1997}. 

Software Engineering (SE) education is not as synonymous with reflective practice as, for example, teacher training, but there is some work on the use of reflection in the teaching of the discipline. There is broad agreement across the existing literature that reflective practice is essential for continuous learning and professional growth in software engineering, e.g.,~\cite{Burden_Steghöfer_2019, BullWhittle}. Reflection allows practitioners to learn from experience, challenge assumptions, and develop new skills~\cite{DybaMaidenGlass} and is often seen as a characteristic of agile software development practices, such as sprint retrospectives~\cite{Babb, pedrosa2021metacognitive}. However, they also point out that in practice, reflection often does not occur as intended in agile projects~\cite{Babb, BullWhittle}.

The literature distinguishes between reflection-in-action (reflecting during an activity) and reflection-on-action (reflecting after an activity)~\cite{DybaMaidenGlass}. Both types are seen as valuable for learning, but there are challenges associated with teaching reflection in a SE context. For example, students may find reflection difficult or intimidating~\cite{Burden_Steghöfer_2019}, and academic settings do not always provide authentic contexts for reflection~\cite{BullWhittle}. Arguably, however, the work-based contexts described here provide just such authentic experiences on which the students may reflect. It is also notable that several authors discuss a shift towards competency-based education in CS more broadly, with reflection playing a key role in developing and demonstrating competencies~\cite{Burden_Steghöfer_2019, BowersHPCSS23}. This emphasis on reflection underpins the approach taken here, as described below. Finally, many of the authors referred to here also note the lack of long-term empirical studies evaluating the effectiveness of reflective practices in SE education~\cite{BullWhittle, Babb, DybaMaidenGlass}.

So, while there is limited work on the analysis of student reflection in SE education, we draw on examples from other domains wherein reflective reports on workplace learning are assessed. For example, Hatton and Smith define types of reflection and strategies to develop reflective practice in student teachers, analyzing written reports from 60 students over two years~\cite{Hatton_Smith_1995}. Lokker and Jezrawi, meanwhile, examined reflective essays from 95 graduate health informatics students to identify themes and areas for course improvement~\cite{Lokker_Jezrawi_2022}. Our work on SE education builds upon such studies on student reflection from other domains, and places a particular focus on the development of reflective practice over time. Framing such a development narrative required the use of models of reflective practice that offered appropriate coding schemes, and we have chosen to use two complementary models which we discuss next: Boud's model of reflection and the 5R model.

\subsection{Reflection Models for Our Study}

Wong \textit{et al.} previously developed a coding process for analyzing students' reflective reports using Boud \textit{et al}.'s model of reflection~\cite{Boud_Keogh_Walker_2013, Wong_Kember_Chung_CertEd_1995}. They created a coding scheme with six categories corresponding to Boud \textit{et al}.'s elements of the reflective process. For each category, Wong \textit{et al.} developed criteria for identifying that element of reflection in the text:
\begin{enumerate}
    \item {\em Attending to feelings} involves recognising and addressing emotions arising from experiences. This process is crucial because emotions can influence how we perceive and reflect on events. Negative feelings may need to be resolved and positive feelings can be used to reinforce learning. 
    \item {\em Association} focuses on linking new experiences with past knowledge or experiences. It involves making links between what has happened and what is already known.
    \item {\em Integration} involves individuals synthesising new information with existing knowledge, constructing new understanding, or modifying previous one. It represents the deeper learning that comes from reflection, where new insight is integrated into a person's broader knowledge.
    \item {\em Validation} refers to the evaluation of new insights or conclusions that emerged from reflection. It involves assessing the accuracy and reliability of new insights by comparing them with external perspectives or prior knowledge.
    \item {\em Appropriation} is when individuals take ownership of new knowledge or skills, personalising and internalising them. This step involves embedding the learning into one's identity,  ensuring it becomes part of one's skills set for future similar situations. 
    \item {\em Outcome of reflection} focuses on what happens as a result of the reflective process, such as a change in behaviour, an intention to act differently, or an increased readiness to face similar challenges in the future. 
\end{enumerate}
To refine these criteria for each category, multiple coders independently analysed students' reflective reports, identifying paragraphs that provided evidence of the different elements and coding them accordingly. While the authors acknowledged that differentiating between each of the six categories was challenging, the developed criteria provide a basis for coding data against Boud \textit{et al.}'s model. 

Another model for structuring reflection is the 5R framework~\cite{Bain_Ballantyne_Mills_Lester_2002}, which proposes a progressive five-component scale: 
\begin{enumerate}
\item \emph{Reporting} is the lowest scale, where the learner provides a clear, factual description of the context, including the events that occurred, the people involved, and any relevant circumstances. It focuses solely on recounting the experience without any subjective interpretations or analysis. 
\item \emph{Responding} is similarly descriptive, with the learner noting how they responded or reacted to the learning experience. This stage captures emotional and cognitive responses, providing insight into how the learner was affected by the event and why they responded in a certain way. 
\item \emph{Relating} marks a significant shift as the learner moves outwith the specific learning experience, relating the learning to prior knowledge and skills that may have been acquired in different contexts.
\item \emph{Reasoning} steps towards higher abstraction, with the learner trying to make better sense of the learning experience by concentrating on significant factors, and then also taking into account different viewpoints and explanations. 
\item \emph{Reconstructing} is where the learner draws conclusions, and plans for future tasks are drawn or re-drawn based on current reflection. This is the highest level, mapping to what Raelin's model describes as transforming tacit knowledge to explicit knowledge, leading to a change in perspective, the ultimate aim of reflective learning. 
\end{enumerate}
The 5R framework for reflection has previously been used to analyse reflective writing (see, for example,~\cite{Farahian_Avarzamani_Rajabi_2021, Oner_Adadan_2016}) and we followed this precedent, using the criteria for each of the five stages to code the students' reports.

\subsection{Research Question}

Wong \textit{et al}.'s version of the Boud model and the 5R framework, together, offer an established and comprehensive conceptual basis for analysing students' reflective work. Using these two models of reflective learning, our longitudinal study aims to answer the research question: \textit{How does the reflective practice of students on a work-based Software Engineering degree program, or \textit{apprenticeship}, develop over time}? The duration of the program is four years -- in line with traditional undergraduate degree programs across Scotland -- and students spend around 80\% of their time in the workplace, as apprentice Software Engineers. The remainder of their time is spent on campus, taking a range of taught courses in Computing Science. Both the university and the workplace aspects of the program are assessed, the latter through a combination of reports, presentations, and portfolios. Further information on the design of the program may be found in \cite{BarrP19, Nabi_Andrei_et_al_2025}.

Here, we offer analyses of a student cohort's reflective practice over the four years of the program, first using Boud \textit{et al.}'s Model of Reflective Practice, then Bain \textit{et al.}'s 5R Framework for Reflection. 

\section{Methodology}

This qualitative study employed a thematic analysis approach set out by Braun and Clarke~\cite{Braun_Clarke_2006} with \textit{a priori} set of codes drawn from the above two reflection models, Boud {\em et al.}'s model and Bain {\em et al.}'s 5R model, to facilitate a structured approach to deductive thematic analysis of the students' reflective reports. We use thematic analysis as it enables a deep exploration of themes in written reflective reports, helping us to uncover how students internalise learning through workplace experiences. 

For this study, we collected reflective reports from students across four years to analyse their development of competencies through workplace learning. In Years 1 and 2, students submit a 750-word reflective report, focusing on how they applied academic knowledge in the workplace and developed key meta-skills. These reports are assessed based on knowledge comprehension, quality of reflection, and academic communication. In Years 3 and 4, students submit a longer reflective report (1500–2000 words). These latter reports emphasise higher-level reflection on their progress toward explicit and implicit competencies, using a competency framework for reference. This type of assessment focuses on reflection quality and clarity of communication, and contributes 10\% of the final course grade in each year. 
The reflective reports across all years are expected to be well-structured and professionally presented, offering valuable insights into students' evolving technical skills, knowledge, and professional dispositions. These reports provide key qualitative data for understanding the students' reflective learning and growth.

The number of student reports included in the analyses varies slightly year-to-year, as students drop out or repeat years, while an additional student joined the cohort in Year 2. Consequently, the sample sizes for each year are as follows: Year 1, \textit{n} = 36; Year 2, \textit{n} = 38 (includes one repeating student); Year 3, \textit{n} = 38; and Year 4, \textit{n} = 37. To ensure participant anonymity, we have adopted a coding system where each student is represented by an identifier in the format S\#/Y\#, with 'S\#' denoting the student number, while 'Y\#' stands for the academic year. For example, S16/Y1 refers to Student 16 in Year 1. This system ensures confidentiality while allowing for the tracking of individual progress across different years.

After familiarising themselves with the data, one author coded the complete dataset against both models, collecting examples of each aspect or stage of the relevant model, where present, in a spreadsheet. Following discussion with the other authors, analysis using the 5R model was refined to allow for `minimal' evidence of a particular stage. Thus, for every student assignment, each stage was deemed to be present, minimal, or not present (absent). This approach allowed for a more nuanced analysis that captured instances where students partially demonstrated that they had achieved a particular stage of reflection (i.e., minimal). The authors agreed that coding using the Boud \textit{et al.} model did not necessitate the inclusion of a `minimal' option; as reported by Wong \textit{et al.} the challenge here was in distinguishing between the elements of the model, not the extent to which the elements were present. To ensure trustworthiness of the analysis, the two other authors then each carried out analysis of a portion of the data, against both models, with disagreements discussed by all three authors and resolved.

Ethical approval to use the students' reports was obtained from the College Ethics Committee as part of a larger project. The students consented to their reflective reports being used for research purposes and we adhered to privacy protocols. As the reports describe work carried out within commercial companies, confidentiality must be observed and the data may not be shared, beyond that reproduced here.

\section{Results \& Discussion}
The results below are presented in relation to each of the selected models of reflection: first, Boud \textit{et al.}'s model, then Bain \textit{et al.}'s 5R model. 

\subsection{The Model of Reflective Process} 
\begin{figure*}
\centering
\includegraphics[width=0.8\textwidth]{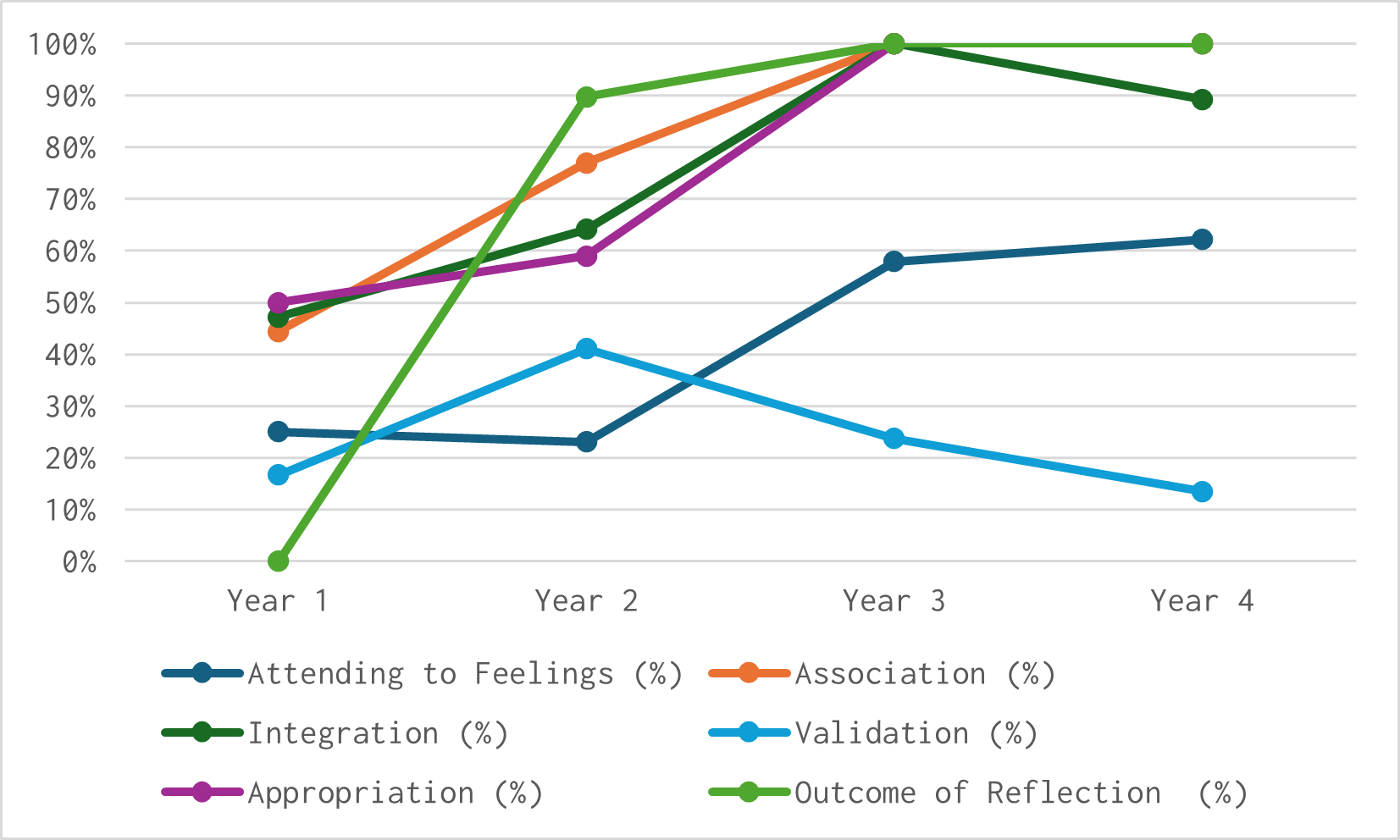}
\caption{The percentage of student reports that contained evidence of each element of Boud \textit{et al.}'s Model of Reflective Process, plotted over time.}
\label{fig:graphBoud}
\end{figure*}

The data relating to each element of Boud \textit{et al.}'s model, as identified in the students' reports, is discussed below. Figure~\ref{fig:graphBoud} illustrates how the prevalence of each element developed over time, with the average number of elements identified per student increasing from Year 1 (\textit{Mo} = 2) to Year 2 (\textit{Mo} = 4) to Years 3 \& 4 (\textit{Mo} = 5).

\subsubsection{Attending to feelings}
While only a quarter of the students provided evidence of attending to their feelings in their first year, it is notable that negative feelings are so prevalent at this early stage, including apprehension, self-doubt, and feeling overwhelmed. One student goes so far as to describe feelings of terror:

\begin{quote}
    \textit{Being accepted into} [the company] \textit{was both exhilarating and terrifying, especially being employed by such a highly sought after and experienced company. In the beginning of my journey I relied on university to provide the fundamentals for starting my subject. As it all felt so vast I realised early on that it was important to pick courses that were most usable/beneficial for getting started.} -- S16/Y1
\end{quote}

This form of reflection was certainly more prevalent in Years 3 and 4, but did not reach the levels of prevalence observed for other forms, with around 60\% of students reflecting on their feelings across both of their final years. An increase in more positive feelings may also be observed, for example:

\begin{quote}
    \textit{As someone who has never enjoyed public speaking or been good at team working, upon reflection it is clear that working at} [the company] \textit{and attending university has greatly improved my confidence.} -- S27/Y3
\end{quote}

\begin{quote}
    \textit{Previously incidents like this would have definitely angered me but recently I have adopted yet again a more relaxed perspective. As long as I test what I am given to the specified requirements, then situations like this are fully at the hands of the} [business analyst] \textit{and I can only do my best to ensure that I hold up from my side.} -- S33/Y3
\end{quote}

\subsubsection{Association}
The ability to associate -- to connect new situations to existing knowledge -- was more prevalent from the outset of the program, with the proportion of students demonstrating this form of reflection rising from 44\% in Year 1 to 100\% in both Year 3 and Year 4. Early examples, such as that below, connect taught material with workplace experience:

\begin{quote}
    \textit{One of the most useful skills that I learned from my first-year modules is the ability to program Web applications. As part of my first major project with my assigned team, I was to help create a web page with a set of features that needed to be implemented. While the programming language used was different from what my team's project used, the module's lab helped me greatly in understanding what the process of creating a web app entails.} -- S12/Y1
\end{quote}

Students continue to associate university material with the reality of the workplace in later years, too:

\begin{quote}
    \textit{I had previously learned about the scrum framework in university and having knowledge of how scrum works before actually being involved with it was very helpful in being prepared for what would be expected of me in a scrum team.} -- S34/Y3
\end{quote}

However, it was not just the taught theory that helped prepare students for the workplace:

\begin{quote}
    \textit{University gave me various concurring projects and deadlines to meet whilst also balancing lectures with regards to knowledge gathering for my courses. This taught me to plan-ahead my pieces of work to avoid deadline penalties and mirrors very nicely into scenarios at work.} -- S36/Y3
\end{quote}

\subsubsection{Integration}
Integration -- synthesising new insights gained from reflection with existing knowledge -- is closely related to association, and follows a similar increase in prevalence across the four years of the program. One early example involves the integration of design pattern knowledge developed at university:

\begin{quote}
    \textit{The university material also covered the use of design patterns to reduce dependencies between modules and separate the code into distinct, logical groups. My team heavily valued the use of design patterns, especially as our project was built upon the .NET framework which relied on the MVC pattern.} -- S32/Y1
\end{quote}

In this Year 3 example, the student reflects explicitly on the integration of the university environment and the workplace environment:

\begin{quote}
    \textit{It is clear to me that the two environments complement each other, creating different experiences that have moulded my professional character. This has resulted in a balanced set of dispositions being developed allowing me to critically analyse problems and translate them into production ready solutions.} -- S42/Y3
\end{quote}

There are also signs of what might be termed \textit{meta-reflection}, as some students begin to reflect on the reflective practice that their apprenticeship has required of them:

\begin{quote}
    \textit{Through the completion of numerous different coursework throughout my time at university and being able to read the feedback provided with each on has allowed for me to develop a certain degree of self-reflection when it comes to any piece of work I create.} -- S35/Y4
\end{quote}

\subsubsection{Validation}
Validation goes a little further, requiring students to determine the relevance and authenticity of the ideas to which they have been exposed. As seen in Fig.~\ref{fig:graphBoud}, this was one aspect of reflection that actually decreased in prevalence over the four years. However, examples may be found, particularly in the Year 1 and Year 2 data. In the example below, the student reflects critically (but positively) on the length of sprints in their workplace:

\begin{quote}
    \textit{My team runs two-week agile sprints as some tasks require quite a bit of time to work on, so if we had shorter sprints more tasks would need to be carried over, so it's just the right length for our team to run efficiently.} -- S10/Y1
\end{quote}

In another example, a student suggests that testing of the software was less relevant in their particular context:

\begin{quote}
    \textit{As }[the company] \textit{are a start-up company the development time was extremely limited. This meant the focus was placed on coding to the requirements, reviewing the design and fixing if needed. Due to the time constraints testing extensively was pushed until after beta launch.} -- S41/Y2
\end{quote}

In both these examples, the students are drawing on what they have learned at university and making a judgement on the validity of those ideas within the context of their workplace. In the former case, the student sees the typical two-week sprints introduced at university as being appropriate to their workplace; in the latter, the emphasis on extensive testing is seen to be less valid within the context of a start-up.

\subsubsection{Appropriation}
Appropriation is about the students making their new-found knowledge their own, and this was another form of reflection that increased over the course of the program. One early example:

\begin{quote}
    \textit{Over time, I have become more proficient with both packages as I continue to work with them daily. By doing so, I have been able to slowly but surely improve my code in order to make it more efficient.} -- S39/Y2
\end{quote}

Later examples see students reflecting on how they have appropriated the tools and techniques to which they have been exposed, for example:

\begin{quote}
    \textit{Even with the presence of a more efficient IDE debugger, I still found myself reverting to old habits and using print statements instead. However, after reminding myself to use the debugger, I started to implicitly learn how to use it properly, and I kept getting better at it over time, so it is now my desired tool when it comes to debugging.} -- S28/Y3
\end{quote}

\begin{quote}
    \textit{By being given the tools to be able to research and analyse new information properly by university and being put into }[a] \textit{scenario in the workplace which allows me to implement the things I have learned and researched, has allowed me to take full advantage of any emerging technologies and integrate them into my software engineering practices.} -- S35/Y4
\end{quote}

\subsubsection{Outcome of Reflection}		
Evidence of outcomes of reflection such as new perspectives or changes in behaviour were almost entirely absent in the Year 1 data, but this form of reflection is present from Year 2 onwards:

\begin{quote}
    \textit{Moving forward, I would like to keep the big-picture view in mind when making technical decisions to lessen the chance of making decisions that sound great short term but are not great or scalable long term.} -- S11/Y2
\end{quote}

\begin{quote}
    \textit{This year in the workplace and university has taught me a lot, I have improved my skills in time/workload management, communication, and interpersonal skills. I have learned to cope with frustration with coworkers through putting myself in their shoes.} -- S33/Y3
\end{quote}

By their final year, students have developed a longer view on their development and, in this case, the student is reflecting on previous reflections, with respect to leadership:

\begin{quote}
    \textit{Looking back, it is clear to me that leadership is a crucial skills for any developer to have even if they are not in a management role, particularly in agile development where developers tend to take a more T-shaped role, spanning many areas of competence.} -- S32/Y4
\end{quote}

In summary, we found that students increasingly evidence some of the `higher' forms of reflection in Boud \textit{et al.}'s model, with Appropriation and Outcome of Reflection becoming more prevalent in the later years of the program. Association and Integration also saw marked increases in prevalence in later years, with Attending to Feelings also increasing over time, but not to the same extent as the aforementioned forms of reflection. The outlier is Validation -- wherein students assess the authenticity their experience -- which tails off in later years. Further study would be required to explore why this is the case, but a possible explanation may be that the students have become more accepting -- and less critical -- of their apprenticeship experience over time.

\subsection{The 5R Framework for Reflection}
\begin{figure*}
\centering
\includegraphics[width=0.9\textwidth]{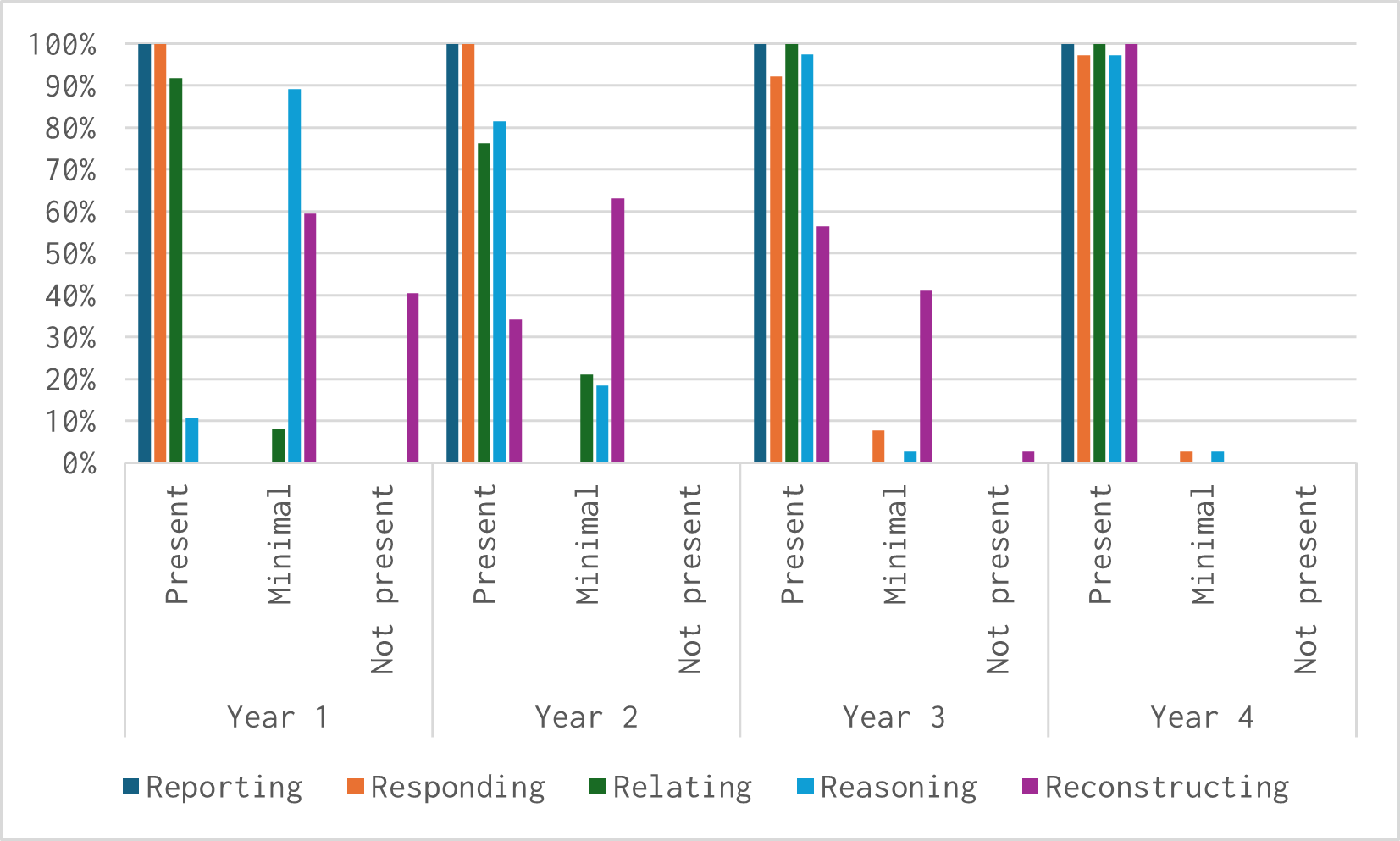}
\caption{The prevalence of each stage of the 5R model of reflection, as identified in reports submitted by students across four years of the program.}
\label{fig:5R}
\end{figure*}

In this section, each of the 5R stages is considered in turn, with the more advanced stages (Relating, Reasoning, and Reconstructing) afforded more space for discussion. As noted above, our coding allowed for the identification of `minimal' evidence for a particular stage: Figure~\ref{fig:5R}, in particular, illustrates how the students' reports typically move from exhibiting minimal or no evidence of a given stage, to evidence of that stage being fully present.

\subsubsection{Reporting}
As might be expected, this most fundamental stage of reflection identified by Bain \textit{et al}. was evident in every student's report, across every year of the apprenticeship program. Examples of students reporting on the context of their work include: 
\begin{quote}
\textit{At university I was given my first true exposure to the power of automated testing.} -- S17/Y1 
\end{quote}
\begin{quote}
\textit{In order to maintain} [the application], \textit{I was given a ticket to decommission JBoss and have the packages removed.} -- S20/Y2 
\end{quote}
\begin{quote}
\textit{During the past year, I have been working with two different teams at my workplace.} --  S24/Y4
\end{quote}

\subsubsection{Responding}
Providing observations, feelings, and thoughts on their experience was similarly prevalent across all four years. For example:
\begin{quote}
\textit{I was impressed at the versatility of automated testing.}  -- S1/Y1 
\end{quote}
\begin{quote}
\textit{I definitely felt a sense of accomplishment at going out of my comfort zone in taking in a technically challenging and involved series of tickets.} -- S39/Y2
\end{quote}
\begin{quote}
\textit{Even though I learnt about user points at university, the challenge I found the most difficult was applying the correct number of points to the stories}'' -- S29/Y4
\end{quote}
However, in a small number of cases, the evidence for the Responding stage of reflection was coded as `minimal', where the student is doing little more than reporting on the context of their work. For example: 
\begin{quote}
    \textit{I remember being assigned the first Jira and thinking I have no idea how to generate automated email.} -- S17/Y3
\end{quote}

\subsubsection{Relating}
The evidence for students relating their experience to existing knowledge and skills was strong from the outset of the program, but strongest in Years 3 and 4. In the first year of the program, the students referred to their experience prior to the program, as well as their initial university classes, for example: 
\begin{quote}
    \textit{Previously I have studied film-making and worked in a supermarket, so software development was a big shock to the system for me.} -- S5/Y1
\end{quote}    
In Year 2, some of the examples of this stage of reflection were again coded as `minimal', as only brief connections between workplace experiences and university learning were made. However, most students were able to relate their learning to the workplace, sometimes purposefully so: 
\begin{quote}
\textit{I took it upon myself to design an application for this using the knowledge I had gained from my first year in university.} -- S40/Y2
\end{quote}

By Year 3, students are well able to relate their university studies -- which, by this stage, have covered most of the basics of Software Engineering -- to their workplace activities. As one student puts it: 
\begin{quote}
\textit{My university courses gave a solid theoretical basis in software engineering. They have taught me basic concepts, programming languages, design principles, and software development processes.} -- S21/Y3
\end{quote}
Other students reflect on the technical skills gained: 
\begin{quote}
\textit{Another example of the knowledge used every day would be that I use BitBucket during my work which we use as a tool for our version control. It is equivalent to GitHub which we used at the university}. -- S24/Y3
\end{quote}
Still others cite professional skills, such as presenting: 
\begin{quote}
\textit{Throughout university, I completed presentations, which taught me important skills such as being able to articulate myself concisely and clearly, as well as design both visually appealing as well as functional documents to relay knowledge to my audience.} -- S28/Y3
\end{quote}

The students continue to relate their workplace experience to the technical and professional skills learned at university throughout their Year 4 reports. However, some additional reflections, which do not fall into this category, may be highlighted. For example, one student reflects not on what they were taught at university, but on how they compared to other students -- specifically, those enrolled on the traditional degree program, as opposed to the apprenticeship: \begin{quote}
\textit{When I compare myself to my university peers, I think that I possess more self-assurance, particularly when it comes to presenting or speaking up about ideas in group projects.} -- S46/Y4
\end{quote}
Others reflect on their personal growth, which may be related to earlier workplace experience, rather than their university experience:
\begin{quote}
    \textit{In the previous three years I have gotten feedback that I am too polite to the point where I didn't want to chase issue however on the last set of feedback from my line manager this was no longer noted to being an issue.} -- S25/Y4
\end{quote}

\subsubsection{Reasoning}
The evidence for students' ability to reason about their experiences develops across the four years of the program, with notable increases in such evidence between Years 1 and 2, and between Years 2 and 3. In the early years, much of the evidence for reasoning comes in the form of analysing the effectiveness of a solution they have developed in the workplace. In later years, students begin to reason about a broader range of factors, often relating to the complementary (or contrasting) nature of the university and workplace experience. For example: \begin{quote}
\textit{I feel practical experience with a task or a programming language with an actual programme is much more helpful for gaining an understanding for how it works in a real-world situation than with lecture slides and random examples.} -- S34/Y3
\end{quote}
In a related example, a student reflects on what the university environment can -- and cannot -- deliver: 
\begin{quote}
\textit{Experiencing the challenges of agile development in a real-world setting allowed me to see the kinds of unexpected scenarios that developers must face, which the university simply cannot cover.} -- S32/Y4
\end{quote}

This kind of reasoning reflects the WBL ethos and demonstrates the benefit of apprenticeships in higher education: the students' workplace experience can afford opportunities for learning in the `real world' that universities cannot replicate. However, the students here have offered many examples of how they have drawn on their university experience in the workplace, and the two environments are clearly complementary. For instance, one student reasons that being able to learn about processes in a relatively `low stakes' university environment is a strength of the program:

\begin{quote}
    \textit{Learning about software product release practices in an environment that was removed from the constraints of my workplace has been very beneficial. In my work role one person never has responsibility for all construction and deployment of an entire application and its infrastructure due to the complexity of the systems we maintain.} -- S30/Y4.
\end{quote}

There is further evidence of meta-reflection here, too, as students in the later years of the program reflect on the their reflective practice to date, and reason about the wider implications of such practice:

\begin{quote}
    \textit{The reflective process itself, prompted by the need to write this essay, has been a valuable exercise in recognising and articulating the competencies developed over the course of my apprenticeship. It has enabled a deeper understanding of how each experience, whether in the workplace or through academic pursuits, contributes to the development of a comprehensive skill set.} -- S42/Y4
\end{quote}

\subsubsection{Reconstructing}
The final stage of the 5R model involves drawing conclusions about future practice based on the reconstruction of previous experience. As with Reasoning, the evidence for this stage increases across the four years of the program, in an almost linear fashion. 

In Year 1, no strong evidence of Reconstructing was identified in any of the students' reports, although many reports contained examples that were coded as `minimal'. For example, this student suggests that they may be able to draw conclusions about their practice \textit{after} they have gained some specific experience: 

\begin{quote}
\textit{I hope that by shadowing the development team, I will be able to directly see these development concepts used in practice within a business environment.} -- S3/Y1
\end{quote}

By Year 2, students provided much stronger evidence for having achieved this stage of reflection, outlining lessons learned about efficiency, clean code, test-driven development, and more, upon which they would subsequently act. For example:

\begin{quote}
    \textit{The benefits of testing are not realised immediately and so, the task of creating effective tests is postponed until it becomes tech debt. This can be remedied through the adoption of test-driven development, whereby business analysts work with developers when creating user stories to create the test cases before coding commences.} -- S42/Y2
\end{quote}

By Year 3, more than half of the students were clearly engaged in reconstructing their experience and looking to the future. In particular, students identified the need to develop certain skills and competencies that, on reflection, they felt they lacked. These ranged from developing technical competencies to improving interpersonal skills. By Year 4, all of the students were demonstrating this form of reflection, with each of the reports featuring some form of future planning, based on the students' experience to date. For example: 
\begin{quote}
\textit{Given my current experience as a software developer the next sensible step for me is to aim to become a line manager in the near future.} -- S46/Y4 
\end{quote}
\begin{quote}
 \textit{To improve the standardisation of processes going forward, I will document expected system behaviours and defect channels on Confluence before handing over to a new team.} -- S14/Y4 
\end{quote}
\begin{quote}
\textit{In the future, I would love to put this knowledge into practice and use TDD in my future tasks.} -- S29/Y4
\end{quote}

Finally, this student outlines a general plan for the future, building on what he has learned, and how he has learned it:

\begin{quote}
    \textit{In the future, I would like to take ownership of similarly large features to further apply the lessons I have learned, as I have come to see that, in regards to development methodologies and practices, it is often more valuable to learn them through application as opposed to simply relying on theoretical knowledge, which may not always reflect the nuances of the real world.} -- S32/Y4
\end{quote}

In summary, we found that Reporting and Responding stages were consistently present across all student reports, showing clear descriptions and personal reactions to workplace experiences. Relating becomes more evident in later years as students connect their workplace activities with knowledge acquired at the university. Reasoning and Reconstructing showed a significant development over time, with Year 3 and Year 4 students demonstrating deeper critical thinking and future planing based on their experiences, which highlights the benefits of WBL in fostering reflective practices. 

\section{Conclusion}

The work presented here has certain limitations. As noted by Wong \textit{et al.}, using their criteria to code against the Boud \textit{et al.} model is challenging, with the finer distinctions between elements of the model often difficult to make. However, we were satisfied with the results obtained here, which revealed clear trends in the evolution of our students' reflective practice. Through discussion and refinement of the coding process, we are also confident in our analysis of the data with respect to the 5R model. 

Coding against both models provides an opportunity for validation, and it is notable that patterns in the development of related aspects of the two models are reflected in our analysis of the data. For example, the increased prevalence of what Wong \textit{et al}. term `Outcome of Reflection' in the Boud \textit{et al.} model follows a very similar pattern to that for the closely-related `Reconstructing' stage of the 5R model.

It is also important to note that this was not an experimental study: we analysed the reports submitted by our students, and make no claims about causation. A strength of the work is its longitudinal nature, with data gathered from the same cohort of students over four years; however, these data were not collected in a uniform fashion. The assigned reports on which the work is based are \textit{similar} in nature, with each asking the students to reflect on what they have learned as an apprentice, but no attempt was made to standardise these assignments. What remains constant is the \textit{context} in which they were completed, i.e., as part of each year's work-based learning module. Future work might standardise the assignments across all four years, although care must be taken to ensure that the guidance given to students did not overly influence the nature of the reflection, for example, to fit with the models used for analysis. Further work might also seek to compare the development of apprentices' reflection with that of their traditional student peers. 

Bearing in mind these limitations, the work provides insight into how the reflective practice of students on a work-based Software Engineering program develops over time. Based on the evidence presented here, it is apparent that these students move increasingly towards more sophisticated forms of reflection as they progress and, given the importance of reflection in our students' learning, this is a positive outcome. The development of `higher' forms of reflection was expected -- the intention of the work-based learning modules is to create reflective life-long learners~\cite{UofG} -- and our analysis appears to confirm this expectation.

Since traditional Computing Science and Software Engineering degree programs -- i.e., those that are not work-based in nature -- typically place less emphasis on reflective practice, perhaps it is worth considering how we might embed more reflection into our campus-based degree programs. Indeed, the approach outlined here -- using the Boud \textit{et al}. and Bain \textit{et al.} models to analyse students' reflective writing -- may be applied in the context of placements or internships. Certainly, as the glimpses of meta-reflection observed here suggest, our students are aware of the pedagogical value of reflecting on their practice.

\bibliographystyle{IEEEtran}
\bibliography{_bibliography}

\vspace{12pt}

\end{document}